\documentstyle [12pt,epsfig,aaspp4] {article}

\newcommand{\be}{\begin{eqnarray}}
\newcommand{\ee}{\end{eqnarray}}
\def\lsim{\mathrel{\rlap{\lower3pt\hbox{\hskip1pt$\sim$}}
    \raise1pt\hbox{$<$}}} 
\def\gsim{\mathrel{\rlap{\lower3pt\hbox{\hskip1pt$\sim$}}
    \raise1pt\hbox{$>$}}} 
\newcommand{\msun}{\mbox{~$M_\odot$}}

\newcommand{\nscoc}{(ns, co)_{\cal C}}
\newcommand{\nscoe}{(ns, co)_{\cal E}}

\newcommand{\lmbh}{\mbox{$lmbh$}}

\lefthead{Brown, Lee, Portegies Zwart, \& Bethe}
\righthead{Evolution of $(ns,co)$ binaries}

\begin{document}


\title{Evolution of Neutron-Star, Carbon-Oxygen White-Dwarf Binaries}
\author{G. E. Brown, C.-H. Lee}
\affil{Department of Physics \& Astronomy,
        State University of New York,
        Stony Brook, New York 11794, USA \\
      Korea Institute for Advanced Study, Seoul 130-012, Korea}
\author{S.\, F.\, Portegies Zwart\altaffilmark{1}}
\affil{Department of Astronomy,
       Boston University, 725 Commonwealth Avenue,
       Boston, MA 01581, USA}
\author{and H.A. Bethe}
\affil{Floyd R. Newman Laboratory of Nuclear Studies,
       Cornell University, Ithaca, New York 14853, USA}

\altaffiltext{1}{SPZ is Hubble Fellow}

\begin{abstract}
At least one, but more likely two or more, eccentric
neutron-star, carbon-oxygen white-dwarf binaries with an
unrecycled pulsar have been observed. According to the standard
scenario for evolving neutron stars which are recycled in common
envelope evolution we expect to observe $\gsim 50$ such circular
neutron star-carbon oxygen white dwarf binaries, since their formation
rate is roughly equal to that of the eccentric binaries and the
time over which they can be observed is two orders of magnitude
longer, as we shall outline.
We observe at most one or two such circular binaries
and from that we conclude that the standard
scenario must be revised. 

Introducing hypercritical accretion into
common envelope evolution removes the discrepancy by converting the
neutron star into a black hole which does not emit radio waves, and therefore
would not be observed.
\end{abstract}

\section{Introduction}
\label{sec0}

We consider the evolution of neutron-star, carbon-oxygen white-dwarf
binaries using both the Bethe \& Brown (1998) schematic analytic
evolutions and the Portegies Zwart \& Yungelson (1998) numerical
population syntheses.

The scenario in which the circular neutron-star, carbon-oxygen
white-dwarf binaries (which we denote as $\nscoc$ hereafter)
have gone through common envelope evolution is
considered. In conventional common envelope evolution for the circular
binaries it is easy to see that the observed ratio of these to
eccentric binaries (hereafter $\nscoe$) 
should be $\sim 50$ because: (i) The formation rate
of the two types of binaries is, within a factor 2, the same.  (ii)
The magnetic fields in the circular binaries will be brought down by a
factor of $\sim 100$ by He accretion in the neutron-star, He-star
phase following common envelope evolution just as the inferred pulsar
magnetic field strengths in the double neutron star binaries are
brought down (Brown 1995).  In the eccentric binaries the neutron
star is formed last, after the white dwarf, so there is nothing to circularize
its orbit. More important, its magnetic field will behave like that of
a single star and will not be brought down from the $B\sim 10^{12}$ G
with which it is born. (At least empirically, neutron star magnetic fields
are brought down only in binaries, by accreting matter from the companion star,
Taam \& Van den Heuvel 1986, although Wijers 1997 shows the situation to
be more complex.) Neutron stars with higher magnetic fields can
be observed only for shorter times, because of more rapid spin down from
magnetic dipole radiation. The time of possible observation goes inversely
with the magnetic field $B$. 
We use the observability premium 
   \be
   \Pi=10^{12}{\rm G}/B
   \ee
(Wettig \& Brown 1996) which gives the relative time a neutron star can be
observed. Given our above point (ii), the circular binaries should have an
observability premium $\Pi\sim 100$ as compared with $\Pi\sim 1$ for the
higher magnetic field neutron star in an eccentric orbit. Correcting for the
factor 2 higher formation rate of the eccentric binaries (point (i) above)
this predicts the factor $\sim 50$ ratio of circular to eccentric binaries.

In our paper we cite one firm eccentric neutron-star, carbon-oxygen 
white-dwarf binary
$\nscoe$ B2303$+$46 and argue for a recently observed second one,
J1141$-$65. 
Portegies Zwart \& Yungelson (1999) suggest PSR 1820$-$11 may also
be in this class, but cannot exclude the possibility that
the neutron star companion is a main sequence star 
(Phinney \& Verbunt 1991).
This would imply that $\gsim 100$ such binaries with circular orbits should
be observed. But, in fact, only one~\footnote{
   We have a special scenario for evolving it; see section~\ref{sec0655}.}
B0655$+$64 is observed if we accept the developing concensus
(Section~\ref{sec4}) that  
those observed $\nscoc$ are evolved with avoidence of common envelope 
evolution.
We are thus confronted by a big
discrepancy, for which we suggest a solution.

In order to understand our solution, we need to review three past works.
In the earlier literature the observed circular $\nscoc$
were evolved through common envelope, e.g., see
Van den Heuvel (1994) and Phinney \& Kulkarni (1994). 
Accretion from the evolving giant progenitor of
the white dwarf was neglected, since it was thought that the accretion
would be held to the Eddington rate of $\dot M_{\rm Edd} \sim
1.5\times 10^{-8} \msun$ yr$^{-1}$, and in the $\sim 1$ year long
common envelope evolution a negligible amount of matter would be accreted.
We term this the standard scenario. However, Chevalier (1993) showed
that once $\dot M$ exceeded $\sim 10^4\dot M_{\rm Edd}$, it was no longer
held up by the radiative pressure due to the X rays from the neutron star,
but that it swept them inwards in an adiabatic inflow. Bethe
\& Brown (1998) employed this hypercritical accretion in their evolution of
double neutron star $(ns,ns)$
and neutron-star, low-mass black-hole binaries $(ns,lmbh)$ and
we shall use the same techniques in binary evolution here. In particular,
these authors found that including hypercritical accretion in the standard
scenario for double-neutron star binary evolution, the first born neutron
star went into a low-mass black hole. 
To avoid the neutron star going through the companion's envelope, 
a new scenario was introduced beginning with a double He star binary.
It gives about the right number of double neutron star binaries.

A new development has been that most of the circular $\nscoc$
are currently evolved with avoidance of common envelope evolution.
In Section~\ref{sec4}, we 
shall summarize this work, carried out independently by
King \& Ritter (1999) and Tauris, Van den Heuvel \& Savonije (2000).
If we accept the new scenario, 
at most one or two circular $\nscoc$ that
went through common envelope evolution have
been observed. Yet, in the standard scenario at least $\sim 50$
of them should be seen.

In this paper we find that in $(ns,co)$ which do go through
common envelope evolution, the neutron star goes into a black hole.
The $(ns,co)$ binaries observed to date have been identified through radio
emission from the neutron star.  Thus, binaries containing a low-mass
black hole would not have been seen.  We discuss masses for which neutron 
stars go into black holes.


Although the main point of our paper relies only on relative formation
rates, we shall show in Appendix A that the Bethe \& Brown (1998)
schematic, analytic analysis agrees well with 
the detailed numerical population synthesis of
Portegies Zwart, once both are normalized to the same supernova rate.

\section{The Problem}
\label{sec1}

\subsection{Standard Scenario vs Observations}
\label{sec_staobs}. 

Portegies Zwart \& Yungelson (1998), in a very careful population synthesis,
have calculated the expected number of newly created binaries of compact stars
(neutron stars or black holes) and white dwarfs.  Among the latter, they
distinguish between those consisting of helium and those 
consisting of carbon-oxygen (denoted $co$). 
To make an eccentric binary containing a neutron star, the supernova
must occur after the carbon-oxygen star has formed (Portegies Zwart \&
Yungelson 1999).
To make a circular binary containing a neutron star it is necessary that its
companion be close so that at some earlier stage in evolution
(but after formation of the neutron star) there was mass transfer
or strong tidal interaction, which requires the companion to
(nearly) fill its Roche Lobe.

Since the Bethe \& Brown (1998) schematic calculations did not include
mass exchange, which is very important in evolving
$\nscoe$ binaries, we need the more complete
calculations of Portegies Zwart \& Yungelson, 
which are listed in Table~\ref{tab2}.  We discuss this in more detail later.
These do not
include hypercritical accretion; i.e., they follow the standard
scenario. In this case the formation ratio of
$\nscoc$ to $\nscoe$ is $17.7/32.1=0.55$.  We now make the case
that if the $\nscoc$ were to be formed through common envelope
evolution (Phinney \& Kulkarni 1994; Van den Heuvel 1994) in the
standard scenario, their
pulsar magnetic fields would be brought down to $B\sim 10^{10}$ G 
because of the similarity to binary neutron star systems in which this
occurs. In detail, this results from helium accretion during the
neutron-star, He-star stage which precedes the final binary (Brown 1995,
Wettig \& Brown 1996).

Detailed calculation of Iben \& Tutukov (1993) for original donor
masses $4-6\msun$ of the white dwarf progenitor show that following
common envelope evolution the remnant stars fill their Roche lobes and
continue to transfer mass to their companion neutron star.  These
remnants consist of a degenerate carbon-oxygen core and an evolving envelope
undergoing helium shell burning. The mass transfer to the neutron star
is at a rate $\dot M < 10^4\dot M_{\rm Edd}$, the lower limit for
hypercritical accretion, so it is limited by Eddington. Van den Heuvel
(1994) estimates that the neutron star accretes about $0.045\msun$ and
$0.024\msun$ in the case of the ZAMS $5\msun$ and $6\msun$ stars, and
$0.014\msun$ for a $4\msun$ star, where these ZAMS masses refer to the
progenitors of the white dwarfs. The accretion here is of the same
order, roughly double,\footnote{
     The He burning time to be used for the progenitor of the white
     dwarf is $\sim 10^6$ years, whereas for the relativistic binary
     pulsars the average time of $5\times 10^5$ years is more
     appropriate, so one would expect a factor $\sim 2$ greater accretion.
     }
the wind accretion used by Wettig \& Brown
(1996) in the evolution of the relativistic binary pulsars B1534$+$12
and B1913$+$16. There the magnetic fields were brought down by a factor
$\sim 100$ from $B\sim 10^{12}$ G to $\sim 10^{10}$ G, increasing the
observability premium $\Pi$ by a factor of $\sim 100$. Thus, the
scenario in which the $\nscoc$ are produced through common envelope
evolution without hypercritical accretion
should furnish them with $\Pi\sim 100$, by helium accretion
following the common envelope. 
Although the detailed description may not be correct, the similarity
of evolution of $\nscoc$ with that of binary neutron stars in the older
works (Phinney \& Kulkarni 1994; Van den Heuvel 1994)
should furnish these with about the same $\Pi$.

There is one confirmed $\nscoe$, namely B2303$+$46, see Table~\ref{tab1},
so there should be about 50 circular ones which went through common
envelope evolution.
Indeed, several circular ones have been observed (see Table~\ref{tab1}),
and one or two of these may have gone through common envelope evolution.
Thus we have a big discrepancy between the standard scenario and the
observations. In the next section, we discuss the possibility
of PSR J1141$-$6545 being $\nscoe$, which enhances the discrepancy.

\subsection{Is PSR J1141$-$6545 $\nscoe$ ?}

\begin{figure}
\centerline{\epsfig{file=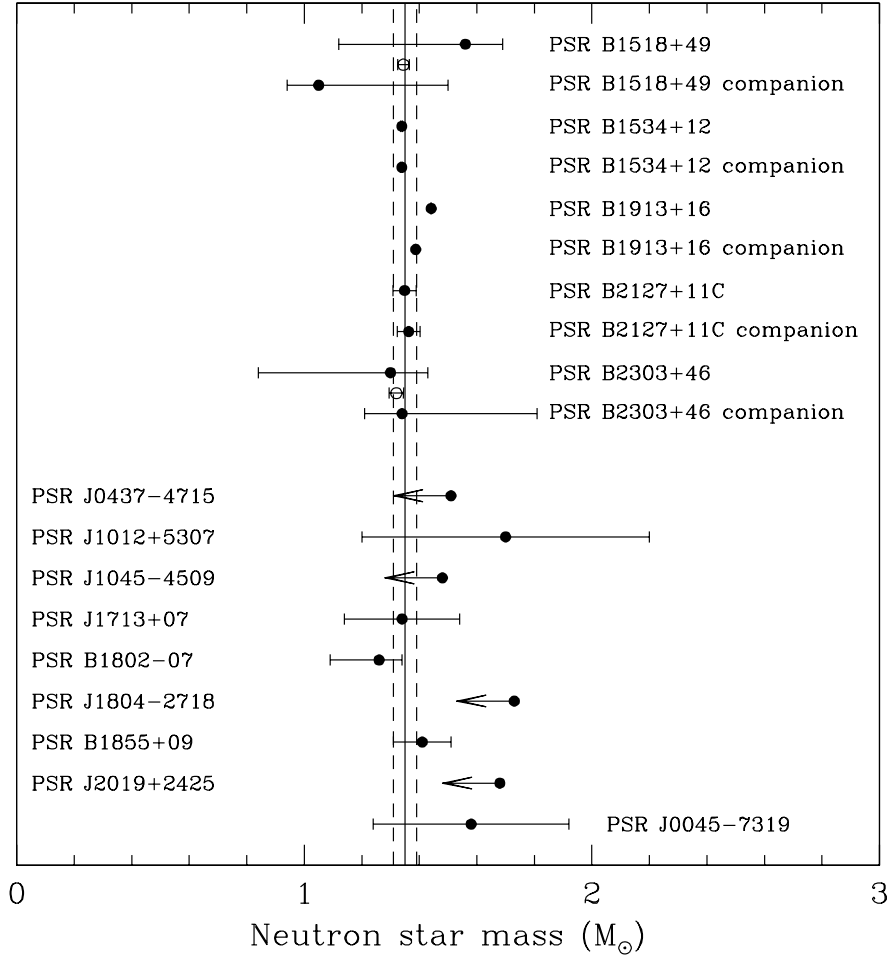,height=5in}}
\caption{Neutron star masses from observations of radio
pulsar system (Thorsett \& Chakrabarty 1999). All error
bars indicate central 68\% confidence limits, except
upper limits are one-sided 95\% confidence limits. Five double neutron
star systems are shown at the top of the diagram. In two cases, the
average neutron star mass in a system is known with much better
accuracy than the individual masses; these average masses are
indicated with open circles. Eight neutron star-white dwarf
binaries are shown in the center of the diagram, and one
neutron star-main sequence star binary is shown at bottom.
Vertical lines are drawn at $m=1.35\pm0.04\msun$. As noted
in our paper, PSR B2303$+$46 has since been shown to have
a white dwarf companion.}
\label{fig1}
\end{figure}

Not only is the eccentric B2303$+$46 quite certain, but a relativistic
counterpart, PSR J1141$-$6545 has recently been observed (Kaspi et al. 2000),
in an eccentric orbit. The inferred magnetic dipole strength is
$1.3\times 10^{12}$ G, and the total mass is $2.300\pm 0.012\msun$.
Kaspi et al. argue that the companion of the neutron star can only be
a white dwarf, or neutron star.
With a total mass of $2.3\msun$, if
J1141$-$65 were to contain two neutron stars, each would have to
have a mass of $\sim 1.15\msun$, well below
the 19 accurately measured neutron star masses, 
see Fig.~\ref{fig1} (Thorsett \& Chakrabarty 1999).

We can understand the absence of binary neutron stars with masses below
$\sim 1.3\msun$, although neutron stars of this mass are expected to
result from the relatively copious main sequence stars of ZAMS mass
$\sim 10-13\msun$ from the argument of Brown (1997). The He stars in
the progenitor He-star, pulsar binary of mass $\lsim 4\msun$ (Habets 1986)
expand substantially during He shell burning. Accretion onto the
nearby pulsar sends it into a black hole. Indeed, with inclusion of
mass loss by helium wind, He stars of masses up to 6 or $7\msun$
expand in this stage (Woosley, Langer \& Weaver 1995). 
Fryer \& Kalogera (1997) find that special kick velocities
need to be selected in order to avoid the evolution of
PSR 1913$+$16 and PSR 1534$+$12 
from going into a black hole by reverse Case C mass transfer
(mass transfer from the evolving He star companion onto the pulsar
in the He-star, neutron-star stage which precedes that of the
binary of compact objects).

Our above argument says that the
first neutron star formed in these would be sent into a black hole
when its companion He star evolved and poured mass on it. Therefore,
we believe the companion in J1141$-$65 must be a white dwarf.
Earlier Tauris \&  Sennels (2000) developed the case that J1141$-$65
was an eccentric neutron-star, white-dwarf binary.
Given the high magnetic field of J1141$-$65 ($1.3\times 10^{12}$ G)
with low observability premium of 0.77, this would increase the
predicted observed number of circular $\nscoc$ which had gone
through common envelope evolution to $\sim 130$ in the
standard scenario.

\subsection{Evolution of Neutron-Star, Carbon-Oxygen White-Dwarf
Binaries with Avoidance of Common Envelope Evolution}
\label{sec4}

Our discussion of the common envelope evolution in the last section applied
to convective donors. In case the donor is radiative or semiconvective,
common envelope evolution can be avoided. Starting from the work of
Savonije (1983), Van den Heuvel (1995) proposed that most low mass
X-ray binaries would evolve through a Her X-1 type scenario, where
the radiative donor, more massive than the neutron star, poured matter
onto its accretion disk at a super Eddington rate, during which time
almost all of the matter was flung off.
This involved Roche Lobe overflow.
Although Van den Heuvel limited the ZAMS mass of the radiative
donor to $2.25\msun$ in order to evolve helium white-dwarf,
neutron star binaries, his scenario has been extended to higher ZAMS
mass donors in order to evolve the carbon-oxygen white-dwarf, neutron
star binaries.
The advection dominated
inflow-outflow solutions (ADIOS) of Blandford \& Begelman (1999) suggest that
the binding energy released at the neutron star can carry away mass,
angular momentum and energy from the gas accreting onto the accretion disk
provided the latter does not cool too much. In this way the binding energy
of gas at the neutron star can carry off $\sim 10^3$ grams of gas
at the accretion disk for each gram accreting onto the neutron star.
King \& Begelman (1999) suggest that
such radiatively-driven outflows allow
the binary to avoid common envelope evolution.

As noted above, for helium white dwarf companions, Van den Heuvel (1995) had
suggested Cyg X-2 as an example following the Her X-1 scenario. King \& Ritter
(1999)
calculated the evolution of Cyg X-2 
in the ADIOS scenario in detail. These authors
also evolved the $\nscoc$ binaries in this way, using donor stars of ZAMS
masses $4-7\msun$. Tauris, Van den Heuvel, \& Savonije (1999) have
carried out similar calculations, with stable mass
transfer. 
These authors find that even for extremely high mass-transfer rates, up to
$\dot M\sim 10^4\dot M_{\rm Edd}$, the system will be able to avoid a
common envelope and spiral-in evolution.

Tauris, van den Heuvel \& Savonije 2000 evolve J1453$-$58, J1435$-$60
and J1756$-$5322, the three lowest entries in our Table~\ref{tab1},
through common envelope. We obtained the eccentricities and
$\dot P$'s for the first two of these (Fernando Camilo, private
communication). The binary J1453$-$58, quite similar to J0621$+$1002
has a substantial eccentricity and clearly should be evolved with a
convective donor as Tauris et al did for J0621$+$1002. The spin
periods of J1435$-$60 and J1756$-$5322 are short, indicating greater
recycling than the other listed pulsars. It would seem difficult to
get the inferred magnetic field down to the $4.7\times 10^8$ G
of J1435$-$60 by the Iben \& Tutukov or Wettig \& Brown accretion
scenarios following common envelope evolution as discussed in
Section~\ref{sec_staobs}.  If, however, one does believe that
J1435$-$60 and J1756$-$5322 have gone through common envelope,
the discrepancy between predicted and observed circular binaries
in the standard scenario is only slightly relieved.


\subsection{Are There Observational Selection Effects ?}

In Table~\ref{tab1b} we have tabulated $S_{400}\times d^2$
in order to see whether the normalized intensity gives
strong selection effects. Note that the 35.95 for B2303$+$46 is
not so different from the 43.56 and 203.35 for B1534$+$12 and
B1913$+$16, respectively. For the circular binaries 
$\nscoc$ the intensities are less, but their empirical Observability
Premium $\Pi$ is much larger. There may be other observational selection
effects, but,
we believe that there are no observational selection effects 
strong enough to compensate for the factor 100  discrepancy
between the observed population and the one expected from the
standard model.
So the problem remains the same.

\section{The Answer}

\subsection{Black Hole Formation in Common Envelope Evolution}
\label{sec2}

We believe the answer to the missing binaries is that the neutron star
goes into a black hole in common envelope evolution, as we now describe.
We label the mass of the neutron star as $M_A$ and that of the
giant progenitor of the white dwarf as $M_B$.
Following Bethe \& Brown (1998) we choose as variables the neutron
star mass $M_A$ and $Y\equiv M_B/a$, where $a$ is the orbital radius.
From their eq.~(5.12) we find
   \be
   \frac{M_{A,f}}{M_{A,i}}=\left(\frac{Y_f}{Y_i}\right)^{c_d-1}
   \label{eq2}
   \ee
where $c_d$ is the drag coefficient.
From Shima et al. (1985) we take
   \be
   c_d = 6.
   \label{eq3}
   \ee
We furnish the energy to remove the hydrogen envelope of the giant B
(multiplied by $\alpha_{ce}^{-1}$, where $\alpha_{ce}$ is the efficiency
of coupling of the orbital motion of the neutron star to the envelope
of B) by the drop in orbital energy of the neutron star; i.e.,
   \be
   \frac{0.6\ G M_{B,i}Y_i}{\alpha_{ce}}
   =\frac 12 G M_{A,i} Y_i \left(\frac{Y_f}{Y_i}\right)^{6/5}.
   \label{eq4}
   \ee
Here the $0.6 GM_{B,i} Y_i$ is just the binding energy of the initial
giant envelope, found by Applegate (1997) to be $0.6 G M_{B,i}^2 a_i^{-1}$,
and the right hand side of the equation is the final gravitational
binding energy
$\frac 12 G M_{A,f} M_{B,f} a_f^{-1}$ in our variables.
Using eqs.~(\ref{eq2}) and (\ref{eq3}) in eq.~(\ref{eq4}) one finds
   \be
   \frac{M_{A,f}}{M_{A,i}} =\left(\frac{1.2 M_{B,i}}{\alpha_{ce} M_{A,i}}
   \right)^{1/c_d}.
   \label{eq5}
   \ee
For the sake of argument, we take the possible range of 
initial neutron star mass to be $1.2-1.5\msun$ (the
upper bound is the Brown \& Bethe (1994) mass at which a neutron star 
goes into a low-mass black hole), and 
the main sequence progenitor masses of 
the carbon-oxygen white dwarf to be $M_{B,i}=2.25-10\msun$.
As we show in Appendix C, in the Bethe \& Brown (1998) schematic model,
mass transfer was assumed to take place when the evolving giant
reached the neutron star, whereas more correctly it begins when the
envelope of the giant comes to its Roche Lobe. For the masses we employ,
main sequence progenitors of the carbon-oxygen white dwarf of $2.25-10 \msun$,
the fractional Roche Lobe radius is
   \be
   r_L\sim 0.5.
   \label{eq7}
   \ee
The binding energy of the progenitor giant at its Roche Lobe is, 
thus, double what it would
be at $a_i$, the separation of giant and neutron star. Therefore, a
Bethe \& Brown $\alpha_{ce}=0.5$ corresponds to a true efficiency 
$\hat \alpha_{ce}\sim 1$, if the latter is defined as the value for which 
the envelope removal energy, at its Roche Lobe, is equal to the drop
in neutron star orbital energy as it moves from $a_i$ to $a_f$. If we
take $\alpha_{ce}=0.5$ in eq.~(\ref{eq5}) we find, given our assumed
possible intervals 
   \be
   1.54 \msun \lsim M_{A,f} \lsim 2.38 \msun.
   \label{eq8}
   \ee
These are above the neutron star mass limit $1.5\msun$ (Brown \& Bethe 1994)
beyond which a neutron star goes into a low-mass black hole.
Thus, all neutron stars with common envelope evolution in our
scenario evolve into black holes. This solves the big discrepancy
between the standard scenario and observation.
The only remaining problem is the evolution of B0655$+$64, which
survived the common envelope evolution, and we suggest a special
scenario for it in the next section.


\subsection{Is B0655$+$64 a problem?}
\label{sec0655}

Van den Heuvel \& Taam (1984) were the first to notice that the
$\nscoc$ system B0655$+$64 might have been formed in a similar
way as the double neutron stars.
The short period
of 1.03 days, magnetic field $\sim 10^{10}$ G,
and the high companion mass of $\sim 1\msun$ make this
binary most similar to a binary neutron star, but with a carbon-oxygen
white-dwarf companion, resulting from probable ZAMS masses
$\sim 5-8\msun$. For a $1.4\msun$ neutron star with $1\msun$ white-dwarf
companion $a_f=5.7 R_\odot$.

The similarity of B0655$+$64 to the close neutron star binaries suggests 
the double helium star scenario (Brown 1995) to calculate the evolution.
The ZAMS mass of the primary  is chosen to be just above the limit  for
going into a neutron star, that of the secondary just below.
For the double He star scenario the
ZAMS masses of primary and secondary cannot be more than $\sim 5\%$
different. 
However, in this case the ratio $q$ of masses is so close to unity
that the secondary will not be rejuvenated (Braun \& Langer 1995:
If the core burning of hydrogen to helium in the companion star is
nearly complete, the accreted matter would have to cross a molecular
weight barrier in order to burn and if $q$ is near unity there is not
time enough to do so.
Thus He cores of both stars will evolve as if 
the progenitors never had more than their initial ZAMS mass.)

What we have learned recently about effects of mass loss (Wellstein \&
Langer 1999) will change the Brown (1995) scenario in detail, but not
in general concept. An $\sim 10\msun$ ZAMS star which loses mass in RLOF
to a lower mass companion will burn helium as a lower-mass star 
due to subsequent mass loss by helium winds, roughly as an $8\msun$ star
(Wellstein \& Langer, in preparation). Thus, the primary must have
ZAMS mass $\gsim 10\msun$ in this case in order to evolve into a neutron
star following mass loss. Although the secondary will not be rejuvenated
as mass is transferred to it, it will burn helium without helium wind
loss because it is clothed with a hydrogen envelope. Thus, a secondary
of ZAMS $8\msun$ will burn He roughly as the primary of $10\msun$ in
the situation considered. 
Given these estimates, a primary of ZAMS mass $M\lsim 10\msun$ will evolve
into a white dwarf, whereas a secondary of mass $\gsim 8\msun$ will end
up as a neutron star. Of course, the former must be more massive than
the latter, but stars in this mass range are copious because this is
the lowest mass range from which neutron stars can be evolved, so
there will be many such cases.

This scenario might not be as special as outlined because the fate of stars of
ZAMS mass $8-10\msun$, which do not form iron cores but do burn in quite
different ways from more massive stars, is somewhat uncertain in the literature.
Whereas it is generally thought that single stars in this range end up as
neutron stars, it has also been suggested that some of them
evolve as  AGB stars ending in white dwarfs.
In terms of these discussions it does not seem unlikely
that with two stars in the binary of roughly the same mass, the first
to evolve will end up as a neutron star and the second as a
white dwarf, especially if the matter transferred in RLOF cannot rejuvenate
the companion.

Van den Heuvel \& Taam (1984) evolved B0655$+$64 by common envelope
evolution.
In taking up the problem again, Tauris, Van den Heuvel \& Savonije (2000)
in agreement with King \& Ritter (1999) find that B0655$+$64 cannot
be satisfactorily evolved with their convective donor scenario. Tauris
et al. suggest a spiral-in phase is the most plausible scenario for the
formation of this system, but we find that the neutron star would go
into a black hole in this scenario, unless the two progenitors burn
He at the same time.

\subsection{Neutron Star Masses}

There is by no means agreement about maximum and minimum neutron star
masses in the literature. The mass determination of Vela X-1 have been
consistently higher than the Brown \& Bethe $1.5\msun$ which is
consistent with well measured neutron star masses in Fig.~\ref{fig1}.
In a recent
careful study at ESO Barziv et al. (2000), as reported by Van
Kerkwijk (2000), obtain
   \be
   M_{NS}=1.87^{+0.23}_{-0.17}\ \msun.
   \ee
Even at 99\% confidence level, $M_{NS}> 1.6\msun$. Taking the
maximum mass to be $1.87 \msun$ and $\alpha_{ce}=0.5$, $(M_{NS})_{min}
=1.2\msun$ one finds from eq.~(\ref{eq5}) that the maximum carbon-oxygen
white dwarf progenitor mass of $\nscoc$ is
   \be
   \left(M_{B,i}\right)_{max}
   =\alpha_{ce} \left(\frac{M_{A,f}}{M_{A,i}}\right)^6
   \frac{M_{A,i}}{1.2} \approx 7.2\msun. 
   \label{eq11}
   \ee
Although there is some uncertainty
in the efficiency $\alpha_{ce}$, the ratio $M_{B,i}/\msun$ is much
more sensitive to $M_{A,f}$ because of the 6th power of the
ratio in eq.~(\ref{eq11}).\footnote{
   With $M_{NS}=1.5\msun$ we get $(M_{B,i})_{max}=1.9\msun$ which is
   below the minimum $M_B$ ($\sim 2.25\msun$) for forming a carbon-oxygen
   white dwarf, so no $\nscoc$ survive the common envelope evolution.}
But then
one cannot explain why no $\nscoc$ (except B0655$+$64) which survived
the common envelope evolution are seen, since this mass is high
enough to give those of the white dwarf companions.

Distortion of the $\sim 20\msun$ B-star companion by the neutron star
in Vela X-1 brings in large corrections
(Zuiderwijk et al. 1977, van Paradijs et al. 1977a, van Paradijs et al.
1997b) making measurement of neutron star masses in high-mass X-ray
binaries much more difficult than those with degenerate companions.

Given the $Y_e\simeq 0.43$ 
at the collapse of the core of a large
star (Aufderheide et al. 1990) one finds the cold Chandrasekhar mass
to be
   \be
   M_{CS}=5.76 \ Y_e^2 \msun \approx 1.06\msun
   \ee
where $Y_e$ is the ratio of the number of electrons to the number of nucleons. 
Thermal corrections increase this a bit, whereas lattice corrections
on the electrons decrease it, so that when all is said and done,
$M_{CS}\gsim 1.1\msun$ (Shapiro \& Teukolsky 1983). 
The major dynamical correction to this is from fallback following
the supernova explosion. 
We believe that fallback in supernova explosions 
will add at least $\sim 0.1\msun$ to the neutron star, 
since bifurcation of
the matter going out and in happens at about 4000 km (Bethe \& Brown 1995).
Thus our lower
limit of $\sim 1.2\msun$ is reasonable.

\section{Discussion and Conclusions}

At least one, but more likely two or more, $\nscoe$ binaries with an
unrecycled pulsar have been observed. According to the standard
scenario for evolving neutron stars which are recycled in a common
envelope evolution we then expect to observe $\gsim 50$ $\nscoc$.
We only observe B0655+64 (which we evolve in our double He-star way)
and possibly one or two binaries that went through common envelope
evolution and from that we conclude
that the standard
scenario must be revised. Introducing hypercritical accretion into
common envelope evolution (Brown 1995; Bethe \& Brown 1998) removes
the discrepancy.

We believe that the evolution of the other $\nscoc$ binaries may
originate from systems with a neutron star with a radiative or
semi-convective companion. The accretion rate in these systems can be
as high as $10^4 \dot M_{\rm Edd}$ but common envelope evolution is
avoided.  This possibility, however, does not affect our conclusion
concerning hypercritical accretion.

It is difficult to see ``fresh" (unrecycled) neutron stars in binaries
because they don't shine for long.  B2303$+$46 (Table~\ref{tab1}) is
the most firm example of a $\nscoe$ binary with a fresh neutron
star.  Although binaries where a ``fresh'' neutron star is accompanied
by a black hole have similar birthrates ($\sim 10^{-4}$
yr$^{-1}$ for both types; Bethe \& Brown, 1999, and Portegies Zwart \&
Yungelson 1999) and lifetime, none are observed.  
In Appendix A we quote results of Ramachandran \& Portegies Zwart
(1998) that show there is an
observational penalty which disfavors the observation of neutron stars with
black holes as companions, because of the difficulty in identifying the
pulsar due to the Doppler shift which smears out the signal in these
short-period objects.
Because of the longer orbital period and lower companion mass of
$\nscoe$, such binaries are less severely plagued by this effect,
although the recently discovered J1141$-$6545 is a relativistic
binary with 5 hr period.
We therefore argue that it is not unreasonable that no $(\lmbh,ns)$ binaries
have yet been observed, but that they should be actively searched for
since the probability of seeing them is not far down from that of
seeing the $\nscoe$'s.

\acknowledgements

We are grateful to Fernando Camilo for information in Table~\ref{tab1}.
We would like to thank Brad Hansen, Marten van Kerkwijk, Ralph Wijers,
Thomas Tauris
and Lev Yungelson for useful
discussions and advice, and thank Justin Holmer for providing us with the
convective envelopes for stars in the giant phase.
GEB would like to thank Brad Hansen also for correspondence which started
him on this problem.
GEB \& CHL were supported by the U.S. Department of Energy under grant
DE-FG02-88ER40388.

SPZ was supported by NASA through Hubble
Fellowship grant HF-01112.01-98A by the Space
Telescope Science Institute, which is operated by the Association of
Universities for Research in Astronomy, Inc., for NASA under contract
NAS\, 5-26555. SPZ is grateful for the hospitality of the State
University of New York at Stony Brook.

\appendix

\renewcommand{\thesection}{A}
\section{Comparison of Population Syntheses}

In this Appendix we first compare results that the Bethe \& Brown (1998)
schematic analytic evolution would have given
without hypercritical accretion with the
Portegies Zwart \& Yungelson (1998, 1999) results of Table~\ref{tab2},
which do not include hypercritical accretion. We can then illustrate how
hypercritical accretion changes the results.

Without hypercritical accretion the $(\lmbh,ns)$ binaries of Bethe \&
Brown would end up rather as neutron star binaries $(ns,ns)$, giving a summed
formation rate of $1.1\times 10^{-4}$ yr$^{-1}$, to compare with
$1.1\times 10^{-4}$ yr$^{-1}$ from the Portegies Zwart
numerical driven population
sythesis results presented in Table~\ref{tab2}.  This good agreement
indicates that kicks that the neutron star receives in formation were
implemented in the same way in the two syntheses.
Introduction of
hypercritical accretion leaves only those neutron stars which do not
go through a common envelope; i.e., those in the double He star
scenario of Brown (1995), with formation rate $10^{-5}$ yr$^{-1}$.
This is much closer to the estimated empirical rate of $8\times
10^{-6}$ yr$^{-1}$ of Van den Heuvel \& Lorimer (1996) which equals
the rate derived independently by Narayan et al. (1991) and Phinney
(1991). Large poorly known factors are introduced in arriving at these
``empirical" figures, so it is useful that our theoretical estimates
end up close to them.  In our theoretical estimates the possibility
described earlier
that the pulsar in the lower mass binary pulsars goes into a black
hole in the He shell burning stage of the progenitor He-star,
neutron-star binary (Brown 1997) 
was not taken
into account and this process may change $\sim$ half of the remaining
neutron star binaries in our evolution into $(\lmbh,ns)$ binaries.

Bethe \& Brown (1998) had a numerical symmetry between high-mass binaries
in which both massive stars go supernova and those in which the more massive
one goes supernova
and the other, below the assumed dividing mass of $10\msun$, did not;
i.e., the number of binaries was equal in the two cases. Taking ZAMS
mass progenitors of $2.3-10\msun$ for carbon-oxygen white dwarfs, we then find
a rate of 
     \be
     R=2 \times \frac{10-2.3}{10} \times 1.1\times 10^{-4}\ {\rm yr}^{-1}
      =16.9 \times 10^{-5}\ {\rm yr}^{-1}
     \ee
for the formation rate of $\nscoc$ binaries. The
$1.1\times 10^{-4}$ yr$^{-1}$ is taken from the last paragraph and 
applies here because of the numerical symmetry mentioned above.
The factor 2
results because there is no final explosion of the white dwarf
to disrupt the binary, as there was above in the formation of the
neutron star. This rate $R$ is to be compared with 
$\nscoc=17.7 \times 10^{-5}$
yr$^{-1}$ in Table~\ref{tab2}.
These $\nscoc$ binaries are just the ones in which the neutron star
goes into a black hole in common envelope evolution, unless the masses
of the two initial progenitors are so close that they burn He at the
same time.
Then a binary such as B0655$+$64 can result, since the two helium stars
then go through a common envelope, rather than the neutron star
and main sequence star.

It is of interest to compare the populations of $(ns,ns)$ binaries
with the $\nscoe$ binaries.
We must rely on the Portegies Zwart result for the latter, which cannot
be evolved without mass transfer, which is not included in the 
Bethe \& Brown evolution. The $(ns,ns)$ binaries involve common
envelope evolution, whereas the $\nscoe$ do not. Thus, results
for the rates should differ substantially in the standard scenario,
which does not include hypercritical accretion, and our scenario which
does. 
The ratio for $(ns,ns)$ and $\nscoe$ from Table~\ref{tab2} are
$10.6\times 10^{-5}$ yr$^{-1}$ and $32.1\times 10^{-5}$ yr$^{-1}$.
The $(ns,ns)$ are recycled in the He-star, pulsar stage by the He wind,
giving an observability premium of $\Pi\sim 100$ (Brown 1995).
The pulsar in the $\nscoe$ is not recycled. Thus, the expected
observational ratio is
     \be
     \frac{(ns,ns)}{\nscoe} \sim \frac{100\times 10.6}{32.1} \sim 33.
     \label{eqC2}
     \ee
Now B2303$+$46, and possibly B1820$-$11 and J1141$-$6545 lie in the
$\nscoe$ class, whereas B1534$+$12 and B1913$+$16 are relativistic
binary neutron stars with recycled pulsars. 
We do not include the neutron star binary 2127$+$11C, although it has
the same $B$ as the other two. It is naturally explained as resulting
from an exchange reaction between a neutron star and a binary which took
place $<10^8$ years ago in the cluster core of M15 (Phinney \& Sigurdsson 1991).
Thus, the empirical
ratio eq.(\ref{eqC2})
is not much different from unity. 
In Bethe \& Brown (1998) common envelope evolution cuts the $(ns,ns)$ rate
down by a factor of 11, only the 1/11 of the binaries which burn He 
at the same time surviving. The remaining factor 3 is much closer to 
observation. 
Furthermore, Ramachandran \& Portegies Zwart  (1998)
point out that there is an observational penalty of a factor of several
disfavoring the relativistic binary neutron stars because of the difficulty
in identifying them due to the Doppler shift which smears out the
signal  in these short-period
objects. 
%
%
%
%
Some observational penalty should, however, also be applied to J1141$-$6545,
which is a relativistic binary. We estimate that the combination of
neutron stars going into black holes in common envelope evolution and
the greater difficulty in seeing them will bring the ratio of 33 
in eq.~(\ref{eqC2}) down to $\sim$ 1 or 2, close to observation.

We have shown that there is remarkable agreement between the Bethe \&
Brown (1998) schematic analytic population synthesis and the computer
driven numerical synthesis of Portegies Zwart \& Yungelson (1998).
This agreement can be understood by the scale invariance in the
assumed logarithmic distribution of binary separations. 
In general we are interested in the fraction of binaries which end up
in a given interval of $a$. E.g., in Bethe \& Brown (1998) that 
fraction was 
   \be
   d\phi =\frac{d(\ln a)}{7}
   \ee
where $d (\ln a)$ was the logarithmic interval between the $a_i$
below which the star in the binary would merge in common envelope
evolution and $a_f$, the largest radius for which they would merge
in a Hubble time. Here 7 was the assumed initial logarithmic
interval over which the binaries were distributed.
Thus, the desired fraction
   \be
   d(\ln a)=\Delta a /a
   \ee
is scale invariant. Mass exchange in the evolution of the binary will
change the values of $a_i$ and $a_f$ delineating the favorable
logarithmic interval, but will not change the favorable $d(\ln a)$.
Of course, when He stars go into neutron stars, the probability
of the binary surviving the kick velocity does depend on the actual
value of $a$, violating the scale invariance. But this does not
seem to be a large effect in the calculations.
In the case of the formation of $\nscoe$ binaries, the neutron star is
formed last, out of the more massive progenitor. Mass transfer is required
for this, because otherwise the more massive progenitor would explode
first. The mass must not only be transferred, but must be accepted, so that
the companion star is rejuvenated (unless $q\sim 1$ as discussed). 
We need the Portegies Zwart \& Yungelson
detailed numerical program for this. In fact, in calculations with this
program (See Table~\ref{tab2}) the formation of $\nscoe$ binaries is
nearly double that of $\nscoc$ ones. 
However, for $q\gsim 0.75$, where $q$ is the mass ratio of original 
progenitors, of the ZAMS
progenitors Braun \& Langer (1995) showed that the transferred hydrogen
has trouble passing the molecular weight barrier in the companion, so
that the latter would not be rejuvenated.
We have not included this effect
here, but roughly estimate that it will lower the predicted numbers of
$\nscoe$ by a factor $> 2$, bringing it down below the number of
 $\nscoc$, exacerbating the problems of the standard model
of binary evolution.

In the literature one sees statements such as ``Population syntheses
are plagued by uncertainties". It is, therefore, important to
show that when the same assumptions about binary evolution are made
and when the syntheses are normalized to the same supernova rates, 
similar results are obtained.
The evolution in the Bethe \& Brown (1998) schematic way is simple, so
that effects in changes in assumptions are easily followed.

\renewcommand{\thesection}{B}
\section{Hypercritical Accretion}

We develop here a simple criterion for the presence of hypercritical
accretion. We further show that if it holds in common envelope evolution
for one separation $a$ of the compact object met by the expanding red
giant or supergiant, it will also hold for other separations and for
other times during the spiral in. We assume the envelope of the giant
to be convective.

In the rest frame of the compact object, Bondi-Hoyle-Lyttleton
accretion of the envelope matter (hydrogen) of density $\rho_\infty$
and velocity $V$ is (for $\Gamma=5/3$ matter)
    \be
    \dot M = 2.23\times 10^{29} (M_{co}/\msun)^2 V_8^{-3}
    \rho_\infty\ {\rm g\ s^{-1}}
    \label{eqA1}
    \ee
where $M_{co}$ is the mass of the compact object, and $V_8$ is the velocity
in units of 1000 km s$^{-1}$, $\rho_\infty$ is given in g cm$^{-3}$.
{}From Brown (1995) the minimum rate for hypercritical accretion is
    \be
    \frac{\dot M_{cr}}{\dot M_{\rm Edd}} = 1.09\times 10^4.
    \label{eqA2}
    \ee
For hydrogen
    \be
    \dot M_{cr}= 0.99\times 10^{22} {\rm g\ s^{-1}}.
    \label{eqA3}
    \ee
Using eqs.(\ref{eqA1}) and (\ref{eqA2}) we obtain
    \be
    (\rho_\infty)_{cr} = 0.44\times 10^{-7} (\msun/M_{co})^2 V_8^3
    \ {\rm g\ cm^{-3}}.
    \ee
Using Kepler for circular orbits
    \be
    V^2=\frac{GM_{\rm tot}}{a}
    \ee
where $M_{\rm tot}$ is the mass of the compact object plus the mass of
the helium core of the companion plus the envelope mass interior to the
orbit of the compact object. One finds
    \be
    (\rho_\infty)_{cr} = 2.1\times 10^{-9}
    (\msun/M_{co})^2 \left(\frac{M_{\rm tot}/10\msun}{a_{12}}\right)^{3/2}
    \ {\rm g\ cm^{-3}}.
    \ee
The $a$-dependence of $(\rho_\infty)_{cr}$ is the same as the asymptotic
density for the $n=3/2$ polytrope which describes the convective envelope.
Thus, if the criterion for hypercritical accretion is satisfied
at one time and at one radius it will tend to be satisfied for other
times and for other radii.
The change of $M_{tot}$ with $a$ is unimportant because from Table~\ref{tab3}
it can be seen that $\rho > (\rho_\infty)_{cr}$ already near the surface
of the star.

In order to check the applicability of hypercritical accretion to the
compact object in the relatively low-mass stars we consider in this
paper, we make application to a $4\msun$ red giant of radius $R=100\
R_\odot$, evolved as pure hydrogen but with inclusion of dissociation
by Justin Holmer (1998).  In the Table~\ref{tab3} we compare the
densities in the outer part of the hydrogen envelope with those needed
for hypercritical accretion.  {}From the table it can be seen that
hypercritical accretion sets in quickly, once the compact object enters
the envelope of the evolving giant.

%
%
%
%

Note that the accretion through most of the envelope will be $>1\msun$
yr$^{-1}$. Since the total mass accreted by the neutron star is $\sim
1\msun$ this gives a dynamical time of $\lsim 1$ year, although the
major part of the accretion takes place in less time. This is in
agreement with the dynamical time found, without inclusion of
accretion, by Terman, Taam \& Hernquist (1995).

\renewcommand{\thesection}{C}
\section{Efficiency}

We discuss the definition of the efficiency of the hydrodynamical
coupling of the orbital motion of the neutron star to the envelope
of the main sequence star.

Van den Heuvel (1994) starts from the Webbink (1984) energetics in
which the gravitational
binding energy of the hydrogen envelope of the giant is taken to be
    \be
    E_{env}=-\frac{G(M_{\rm core}+M_{\rm env})M_{\rm env}}{R}
    \label{eqB1}
    \ee
which results in the envelope gravitational energy
    \be
    E_{env} = -\frac{0.7 GM^2}{R},
    \label{eqB2}
    \ee
where $M=M_{\rm core}+M_{\rm env}$ is the total stellar mass
and the Bethe \& Brown (1998) approximation $M_{\rm core}\simeq 0.3 M$
has been used. 

Applegate (1997) has calculated the binding energy of a convective giant
envelope, obtaining
    \be
    E_B= -0.6 GM^2/R = \frac 12 E_{env}.
    \label{eqB3}
    \ee
Note that $M$ is the total stellar mass, also that $E_B$ is just 1/2 of
the gravitational potential energy, the kinetic energy being included
in $E_B$. 
Eq.~(\ref{eqB3}) was checked independently by Holmer (1998).
Unfortunately, this work was never published.
Van den Heuvel and others have introduced an additional parameter
$\lambda$ that both takes into account the kinetic energy and the
density distribution of the star, $R$ in eqs.~(\ref{eqB1}) \& (\ref{eqB2})
being replaced by $R\lambda$. They use
    \be
    E_B = -\frac{0.7 G M^2}{\lambda R}.
    \ee
With $\lambda=7/6$ this is the same as $E_B$ in eq.~(\ref{eqB3}).
Van den Heuvel (1994) chooses $\lambda=1/2$. The result is
that his efficiency $\eta$ is a factor of 7/3
too high. Thus, his suggested efficiency $\eta=4$ is more like
$\eta\sim 12/7=\hat\alpha_{ce}$.

Bethe \& Brown (1998) used the Applegate result but incorrectly took
the necessary energy to expel the giant envelope as
    \be
    E_g = -0.6 \ GM^2/a_1
    \ee
rather than
    \be
    E_g=-0.6\ GM^2/a_1 r_L,
    \ee
the latter being the correct energy needed to remove the giant envelope
at its Roche Lobe.  The correct efficiency is
    \be
    \hat\alpha_{ce}=(\alpha_{ce})_{BB}/r_L.
    \ee
and since for the binaries considered here with $q\sim 4$ the
fractional Roche Lobe is $r_L\sim 0.5$,
    \be
    \hat\alpha_{ce}\simeq 2 (\alpha_{ce})_{BB} =1,
    \ee
with the Bethe \& Brown (1998) $\alpha_{ce}=0.5$.

Of course, $\hat\alpha_{ce}$ should not vary with Roche Lobe, the
Bethe \& Brown (1998) usage of $\alpha_{ce}$ being in error.

For $\hat\alpha_{ce}=1$, the envelope would be removed from the giant,
but would end up with zero kinetic energy, which is unreasonable.
Thus, without additional energy sources, as discussed earlier in
our note, one would expect $\hat\alpha_{ce}\sim 0.5$, in which case
the kinetic energy of the envelope would remain unchanged in its expulsion.
The Bethe \& Brown (1998) results
were insensitive to changes in $\hat\alpha_{ce}$, which changed the
location but not the magnitude of the favored logarithmic intervals,
as noted by those authors.

In the present case Van den Heuvel's $\hat\alpha_{ce}=1.2$ definitely
indicated the presence of energy sources additional to the drop in
orbital energy, although they are not as large as he indicated.
We have checked that with $\hat\alpha_{ce}=1.2$ and $c_d\gg 1$ in the
Bethe \& Brown (1998) formation we obtain the numerical results of
Table~1 of Van den Heuvel (1994).


\newpage

\renewcommand{\thetable}{1}
\begin{table}
\caption{Simulations, normalized to supernova rate of 0.025 yr$^{-1}$,
assuming 100\% binarity following Case B of Portegies Zwart \&
Yungelson (1998). These simulations do not include hypercritical
accretion.
}
\label{tab2}
$$
\begin{array}{lr} \hline
{\rm binary }       & {\rm model\ B} \\
                &[10^{-5}\,{\rm yr}^{-1}] \\ \hline
(ns, ns)          &  10.6  \\
(bh, ns)          &  1.9  \\
\nscoc        &  17.7  \\
\nscoe        &  32.1  \\ \hline
\end{array}
$$
\end{table}

\def\dd{{\dagger\dagger}}
\def\ss{{\star}}
\def\J{{\rm J}}
\def\B{{\rm B}}
\def\C{{\rm C}}
\def\tmf{{\times 10^{-5}}}
\renewcommand{\thetable}{2}

\begin{table}
\caption{Binary Radio Pulsar Systems : $(ns, ns)$ and $(ns, co)$ binaries.
The Observability Premium $\Pi = [10^{12}\,{\rm G}]/B$.
$M_p$ ($M_c$) means the pulsar (companion) mass, and $f$ the mass function.  }
\label{tab1}
$$
\begin{array}{lccccccccc} \hline
{\rm Pulsar} &P_{\rm orb} &P_{\rm spin}& f     &  M_p  &    M_c   &   e      & d         & B                  & \Pi   \\
             &[{\rm days}]&[{\rm ms}]  &[\msun]&[\msun]&  [\msun] &          &[{\rm kpc}]&   [{\rm G}]        &       \\ \hline
(ns, ns) \\
\J1518+4904   &   8.634  &   40.9  & 0.116  & < 1.75   & > 0.93   & 0.249    & 0.70      & <1.3\times 10^9    & > 769 \\
\B1534+12     &   0.421  &   37.9  &        & 1.339    & 1.339    & 0.274    & 1.1       &  10^{10}           & 100   \\
\B1913+16     &   0.323  &   59.0  &        & 1.441    & 1.387    & 0.617    & 7.13      & 2.3 \times 10^{10} & 43    \\
\B2127+11\C^\dagger & 0.335 & 30.5 &        & 1.349    & 1.363    & 0.681    & 10        & 1.2 \times 10^{10} & 83    \\  \hline
\nscoe \\
\B2303+46     &   12.34  &  1066.  & 0.246  & <1.44    & > 1.20   & 0.658    & 4.35      & 7.9 \times 10^{11} & 1.26  \\
\J1141-6545^\dd & 0.198  &  394.   & 0.177  & < 1.348  & > 0.97   & 0.172    & 3.2       & 1.3 \times 10^{12} &  0.77 \\ \hline
\nscoc^{\star\star} \\
\J2145-0750   &   6.839  &  16.1   & 0.024  &          & 0.515    & 2.1\tmf  & 0.5       & 6\times 10^8       & 1667  \\
\J1022+1001   &   7.805  &  16.5   & 0.083  &          & 0.872    & 9.8\tmf  & 0.6       & 8.4\times 10^8     & 1190  \\
\J1603-7202   &   6.309  &  14.8   & 0.009  &          & 0.346    & < 2\tmf  & 1.6       & 4.6\times 10^8     & 2173  \\
\J0621+1002   &   8.319  &  28.9   & 0.027  &          & 0.540    & 0.00245  & 1.9       & 1.6\times 10^9     & 625   \\
\B0655+64     &   1.029  & 195.7   & 0.071  &          & 0.814    & 0.75\tmf & 0.48^\ss  & 1.26\times 10^{10} & 79    \\ 
\J1810-2005   &   15.01  &  32.8   & 0.0085 &          & 0.34     &          &           & 2.1\times 10^9     & 476   \\
\J1157-5112   &   3.507  &  43.6   & 0.2546 &          & > 1.20   &          &           & < 6.3\times 10^9   & 159  \\
\J1232-6501   &   1.863  &  88.3   & 0.0014 &          & 0.175    &          &           & 9.5\times 10^9     & 105  \\
\J1453-58     &   12.42  &  45.3   & 0.13   &          & 1.07     & 0.0019    &           & 6.1\times 10^9     & 164 \\
\J1435-60     &   1.355  &  9.35   & 0.14   &          & 1.10     & 1\tmf   &           & 4.7\times 10^8     & 2127 \\
\J1756-5322   &   0.453  &  8.87   & 0.0475 &          & 0.683    &          &           &                    &      \\
\hline
\end{array}
$$
$\dagger$: binary in globular cluster M15.
$\dd$: not confirmed yet.
$\ss$: assumed distance.\\
$\ss\ss$ 
The white dwarf mass $M_c$ is calculated assuming 
$M_p=1.4\msun$ and $i=60^\circ$.\\
Refs; Thorsett et al. (1999); B2303: Kerkwijk et al. (1999);
 J2145: Bailes et al. (1994); J1022: Camilo (1995);
 J1603: Lorimer et al. (1996); J0621: Camilo et al. (1996);
 B0655: Kerkwijk et al. (1995); J1141$-$65: Kaspi et al. (2000).
\end{table}

\renewcommand{\thetable}{3}
\begin{table}
\caption{Flux densities at 400 MHz
[Ref: The Pulsar Catalog, Princeton Pulsar Group, http://pulsar.princeton.edu] }
\label{tab1b}
$$
\begin{array}{lrr} \hline
{\rm Pulsar}  & S_{400}       & S_{400} \times d^2        \\
              & [{\rm mJy}]   & [{\rm mJy\cdot kpc^2}] \\
\hline
(ns, ns) \\
B1534+12      & 36.00         &  43.56  \\
B1913+16      &  4.00         & 203.35  \\
B2127+11C     &  0.60         &  56.45  \\
\nscoe \\
B2303+46      &  1.90         &  35.95  \\
\nscoc \\
J2145-0750    & 50.00         &  12.50  \\
J1022+10      & 23.00         &   8.28  \\
B0655+64      &  5.00         &   1.15  \\
\hline
\end{array}
$$
\end{table}

\renewcommand{\thetable}{4}
\begin{table}
\caption{Densities in g cm$^{-3}$for the hydrogen envelope of
a $4\msun$ star of radius $100 R_\odot$. From Holmer (1998).}
\label{tab3}
$$
\begin{array}{ccc}
\hline
r/R_\odot  & \rho  & (\rho_\infty)_{cr} \\ \hline
1          &  0    &                    \\
0.95       & 5.0(-11) & 7.8(-13)        \\
0.90       & 3.5(-9)  & 8.5(-13)        \\
0.85       & 5.8(-8)  & 9.3(-13)        \\
0.80       & 3.9(-7)  & 10.(-13)        \\ \hline
\end{array}
$$
\end{table}

\end{document}